\documentclass[conference]{IEEEtran}
\ifCLASSINFOpdf
\else
\fi
\usepackage{color}                       
\usepackage[all]{xy}
\usepackage{amsmath,amsthm,amsfonts,amssymb,bm} 
\usepackage{graphicx,psfrag}                    
\usepackage[all]{xy}
\usepackage{float}
\newtheorem{prop}{Proposition}[section]
\newtheorem{defi}{Definition}[section]
\newtheorem{exam}{Example}[section]

\begin{document}
%
\title{On the hopping pattern design for D2D discovery with invariant}

\author{\IEEEauthorblockN{Qizhi Zhang, Changqing Yang}
\IEEEauthorblockA{Huawei Technologies Co. Ltd., Beijing, China \\
Email: \{zhangqizhi, changqing.yang\}@huawei.com}

}


%


\maketitle

\begin{abstract}
In this paper, we focus on the hopping pattern design for device-to-device (D2D) discovery. The requirements of hopping pattern is discussed, where the impact of specific system constraints, e.g., frequency hopping, is also taken into consideration. Specifically speaking, we discover and utilize the novel feature of resource hopping, i.e., "hopping invariant" to design four new hopping patterns and analyze their performance. The hopping invariant can be used to deliver information for specific users without extra radio resources, and due to the connection between hopping invariant and resource location, receiver complexity can be significantly reduced. Furthermore, our schemes are designed to be independent of discovery frame number, which makes them more suitable to be implemented in practical systems.

\end{abstract}


%
\IEEEpeerreviewmaketitle

\section{Introduction}

Mobile traffic explosion due to the popularity of smart devices (e.g., smart phones and electronic tablets) and various multimedia applications has imposed a significant impact on the cellular networks. In industry, it is estimated that about 1000 times capacity improvement is necessary in the decade between 2010 and 2020. Although how to meet this requirement is still an open issue, current study shows that device-to-device (D2D) communications have been considered as one of the key techniques in the Third Generation Partnership Project (3GPP) Long Term Evolution Advanced (LTE-Advanced) due to its flexibility in improving local service and offloading data from cellular networks \cite{power0}.

Generally speaking, D2D communication is composed of two relatively independent procedures, i.e., D2D discovery and D2D data communication. In D2D discovery procedure, D2D user equipments (UEs) discover each other with a well design operation flow, while in D2D communication procedure, resource allocation is carried out to facilitate the data transmission. To the best knowledge of the authors, there have been a lot of researches on D2D communication procedure. For example, in \cite{power1}, \cite{power2}, \cite{power3}, power control in D2D communication
 is studied. In \cite{mimo1}, \cite{mimo2}, \cite{mimo3}, D2D communication with Multiple Input Multiple Output (MIMO) is studied.

As the previous procedure in D2D communication, D2D discovery is a key step to determine which users are included in a D2D transmission group. However, the design and optimization of D2D discovery procedure is still a challenge issue and limited work has been done. The main obstruction in D2D discovery is half duplex constraint, which means that a certain D2D UE cannot receive discovery signals from other D2D UEs while transmitting its own, even on different but adjacent frequency bands. Therefore, the radio resources used by a D2D UE to transmit and receive discovery signal should be carefully arranged to enable efficient discovery procedure. In \cite{QC}, a hopping pattern is proposed to facilitate the discovery procedure. In this scheme, periodic discovery frame is designed and each of which consists of $n$ sub frame. In addition, the frequency band of system is divided into $m$
 parallel channels using Single Carrier Frequency Division Multiple Access (SC-FDMA), where $m\leq n$.
\begin{figure}[H]
\centering
\includegraphics[width=0.3\textwidth]{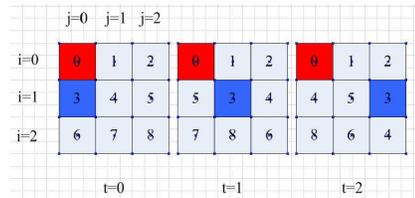}
\caption{Example of hopping pattern}
\end{figure}
 A frequency-time
block shown in Figure 1 corresponds to a unique discovery resource unit and is the basic resource unit for the
device to send or receive a discovery signal. Hence, in one
discovery repetition period, the discovery channel resource is
logically divided into $mn$ discovery resource unit. All devices transmit their discovery signal as following hopping pattern:
\begin{align*}
& i(t)=i(0)+c \mod m \\
& j(t)=(j(0)+i(0)t) \mod n.
\end{align*}
where $i(t)$, $j(t)$ means the time-domain and frequency-domain location the discovery resource used by a certain UE, and $c$ is a integral constant number.

The two equations indicate a certain UE's transmission frequency-time unit $(i(t), j(t))$ in discovery frame $t$,
 which is decided by the frequency time unit $(i(0), j(0))$  in discovery frame $0$. Therefore, the impact of half duplex constraint is weakened. For example, assume UE 0 transmits its discovery signal on red units, i.e., $(i(0), j(0))=(0,0), (i(1), j(1))=(0, 0), \cdots$ while UE 3 transmits its discovery signal on blue units, i.e., $(i(0), j(0))=(1,0), (i(1), j(1))=(1,1), \cdots$. Although they can not receive each other¡¯s discovery signals in discovery frame $0$ as they are transmitting, it is possible for them to receive the signals and find each other in other frames (e.g., frame $1$ or $2$) .

Once hopping pattern is determined, some relation is determined between the discovery resource units in
different discovery frames. To be general, the set of frequency-time units for a certain UE can be regarded as a particular logic discovery resource. For example, we can see the red and blue frequency-time units as the logical discovery resources for UE1 and UE2, respectively.

To the best of the authors' knowledge, few work has been done on the design of hopping pattern for D2D discovery. In \cite{3GPP_QC} and \cite{our_narrow}, new hopping patterns are proposed which are applicable for more general structure of discovery frame. In \cite{3GPP_Huawei}, a pattern "independent of frame number $t$" is proposed which is robust and can reduce overhead if deployed in LTE networks. For more details, please refer to \cite{3GPP_Huawei} or section 2.2 in this paper.

In this paper, we design four new types of hopping pattern, i.e., type A1, A2, B1 and B2, based on a new feature called "hopping invariant", due to which additional information, e.g., the service type, can be easily embedded in the hopping pattern. Besides, since only a small part of the radio resources is connected to a certain hopping invariant, it helps to significantly simplify the operation of the receivers which are only interested in ceratin additional information. Furthermore, we consider the impact of hopping in frequency and the discovery frame number, which makes our scheme more suitable to be implemented in practical system, e.g., LTE.

The structure of this paper is as below: in section 2, we form the problem of discovery with system constraint including hopping invariant, hopping in frequency and irrelevance of discovery frame number. Next we solve the problem with linear algebra theory on finite fields and get our proposed hopping patterns. After analyzing the performance of our schemes, we draw the conclusion in section 4.

Some of the mathematical symbols and expressions used in this paper are listed below:

For a finite set $S$, $|S|$ means the number of elements in $S$.

For two set $X$ and $Y$, $X \times Y$ means the Cartesian product of $X$ and $Y$;
\begin{displaymath}
\begin{array}{rccc}
f:& X & \longrightarrow & Y \\
& x & \mapsto & f(x)
\end{array}
\end{displaymath}
means a map from $X$ to $Y$, which name is $f$.

$\mathbb{Z}$ means the set of integral numbers, $\mathbb{Z}/n\mathbb{Z}$ means the set $\{0, 1, 2, \cdots, n-1\}$ for positive integral number $n$.

\section{Modeling and problem formulation}
We discuss the basic requirement of hopping pattern, together with the constraint of LTE system and the new feature of "hopping invariant". After that, we give the formulation of the problem in subsection \ref{formulation}.

\subsection{Basic requirement for hopping pattern}
\label{basic}

A discovery frame is
divided into some sub frames, we denote the set consists of them by $J$. The frequency band of Physical Uplink Shared Channel (PUSCH) is divided into some
 parallel channels using SC-FDMA, we
denote the set consists of them by $I$. Then the resources in every discovery frame are divided into $|I|\times|J|$ parts, each of which is corresponding to an element in set $I\times J$ and called a
 frequency-time resource in this discovery frame. Such structure allows $|I|\times|J|$ UEs transmit their own
  discovery signal with different frequency-time resources in a single discovery frame. Due to half duplex constraint, if two UEs always transmit discovery signals in same time, they can't receive each other all the time. Hence we have:

{\bf Req 1.} If two UEs transmit their own discovery signal simultaneously in a discovery frame, they should transmit own discovery signal at different times in next discovery frame.

\subsection{Embedding in LTE system}
\label{LTE}
In LTE Frequency Division Duplex (FDD) system, since the uplink load is usually lighter than that of downlink, idle uplink radio resources can be utilized to facilitate the D2D discovery procedure. In the discovery frame, the both ends of band need to be preserved for acknowledgement/non-acknowledge (ACK/NACK) of downlink. Hence if a UE closed to evolved Node B (eNodeb) always transmits own discovery signal on a sub-band closed to an
 end of band, its discovery signal may impose significant interference on ACK/NACK received by eNodeb. (Notice: For the purpose that be discovered by more
 UEs, the power of discovery signal may not be controlled by eNodeb.) To randomize such interference, the frequency coordinate
 of logical discovery resource should be carrier out, e.g., via hopping. Hence we have:

{\bf Req 2.}  The frequency coordinate
 of every logical discovery resource should hop.  \qed

If $(i(t), j(t))$ is totally determined by $(i(0), j(0))$ and $t$, we call
 the hopping pattern deterministic. For example, the hopping pattern defined in section 1.1 is a deterministic hopping pattern. The random hopping pattern is not a deterministic hopping pattern. The deterministic hopping patterns have the
advantage that the receiver can jointly decode the discovery signals from a ceratin UE by combining them in multiple discovery frames.

It easy to see, that for a deterministic hopping pattern $\{(i(t),j(t))\}_{t\in \mathbb{Z}}$, there exists a sequence
of bijections
\begin{align}
\{ f_t: I \times J \longrightarrow I \times J \}_{t \in \mathbb{Z} }
\end{align}
such that $(i(t+1),j(t+1) )=f_t (i(t),j(t))$. We call $\{ f_t \}_{t \in \mathbb{Z}}$ the frequency-time hopping function
of hopping pattern $\{(i(t),j(t))\}_{t\in \mathbb{Z}}$. If there exists a bijection
\begin{align}
f: I \times J \longrightarrow I \times J
\end{align}
such that $f_t=f$ for any $t \in \mathbb{Z}$, we call the frequency-time hopping function
independent of frame number, else call it dependent on frame number.

Consider the next cases:

Case 1: UE1 starts to search D2D discovery signal from discovery frame $t$ , suppose it had detected energy on frequency-time
coordinate $(i(t),j(t) )$, but it can not decoded it. Then UE1 wants to hear it again in next discovery
 period for the purpose of time diversity or joint decoding. Hence UE1 needs to compute $(i(t+1),j(t+1) )$ from $(i(t),j(t) )$.

Case 2: UE1 starts to search D2D discovery signal from discovery frame $t$ , suppose it detected no energy or
rather low energy on frequency-time coordinate $(i(t),j(t) )$, it wants to transmit own discovery signal in
this resource from next discovery frame. Hence UE1 needs to compute $(i(t+1),j(t+1))$ from $(i(t),j(t) )$.

In above two cases, UE1 needs to compute $(i(t+1),j(t+1) )$ from $(i(t),j(t) )$.
 If the frequency-time hopping function is independent of $t$ , UE1 can compute it easily.
 But if the frequency-time hopping function is dependent on $t$ , UE1 needs to know the number $t$
  of current discovery frame. But in several system the system frame number is a number in $\mathbb{Z}/T\mathbb{Z}$.
   For example, in the LTE system, the system frame number is a number in $\mathbb{Z}/1024\mathbb{Z}$.
   For the purpose of decreasing the occupancy ratio of resources,  the interval between two
   discovery frame should be long, hence UE1 can not learn the number $t$
    of current discovery frame from system frame number. Therefore, we have:

{\bf Req 3.} The deterministic hopping pattern with frequency-time hopping function independent on $t$ should be used.

\subsection{Hopping invariant}
\label{invariant}

Discovery signal should carry some information, for example,
 the information about UE identity (ID) and service type. The information can be carried through coding and modulation,
 it can also be carried by frequency-time coordinate. The method to carry information by using
 frequency-time coordinate has the virtue that the information carried by frequency-time coordinate impossibly is decoded wrong.
  Therefore we wish to use the frequency-time coordinate to carry some important information. But the frequency-time coordinate
   vary in every discovery frame. When the hopping pattern $\{(i(t),j(t) ) \}_{ t\in \mathbb{Z}}$ is deterministic,
   a natural idea is to use the frequency-time coordinate $(i(0),j(0) )$ in discovery frame $0$ to carry the information.

For example, when the ¡°information¡± is service type, we can classify $I\times J$ into some non-intersect subsets $U_0,U_1,¡­,U_{q-1}$,
 that corresponding to the service type $0,1,...,q-1$ respectively. Any UE with service type $k$ transmit own discovery signal
 on a discovery resource $(i(t), j(t))$ such that $(i(0),j(0))$ lie in the subset $U_k$.
  If a UE received a discovery signal from a discovery resource $(i(t),j(t))$
   in discovery frame $t$, it needs to compute $(i(0),j(0))$ from $(i(t),j(t))$ and $t$, and get the service type
   of sending UE from $(i(0),j(0))$.

But in several system the system frame number is a number in $\mathbb{Z}/T\mathbb{Z}$.
 For example, in the LTE system, the system frame number is a number in $\mathbb{Z}/4096\mathbb{Z}$.
  For the purpose of decreasing the occupancy ratio of resources of system, the interval between two discovery frame
   should be long, hence receiving UE can not learn the number $t$ of current discovery frame then can not compute $(i(0),j(0))$
    from $(i(t),j(t))$. Then the mechanism to carry information by using $(i(0),j(0) )$ is unsuccessful.

However, in this situation, the invariant of hopping pattern defined as follows can carry information.

\begin{defi}
Let  $\{(i(t),j(t) ) \}_{t \in \mathbb{Z}}$ be a deterministic hopping pattern on
discovery frame structure $(m, n)$. If there exists a set $X$ and a surjection $g: I\times J \longrightarrow X$ such that
 $g((i(t),j(t) ) )=g((i(t'),j(t') ) )$ as maps from $S$ to $I \times J$ for any $t,t' \in \mathbb{Z}$, we call $g$
  an invariant of hopping pattern $\{(i(t),j(t))\}_{t \in \mathbb{Z}}$.
\end{defi}

For example, when the ¡°information¡± is service type, $I \times J$ is classifying by invariant into some non-intersect subsets
$U_0,U_1,¡­,U_{q-1}$ naturally, which invariant equal to $0,1,...,q-1$ respectively, that
corresponding to the service type $0,1,...,q-1$ respectively. Any UE with service type $k$
 transmit own discovery signal on a discovery resource s such that $(i(0),j(0))$ lie in the subset $U_k$. If a UE received
  a discovery signal from a discovery resource s with frequency-time coordinate $(i(t),j(t))$ in discovery frame $t$,
   it compute the invariant from $(i(t),j(t))$, and learn the service type of sending UE.

Moreover, if a UE want to find the UEs with special service type, it only need to decode the discovery signal
in the resources which invariant corresponding to this service type.
Relative to the method decode the discovery signals on all the resources, energy is reduced. Hence we have:

{\bf Req 4.} A pattern with hopping invariant should be used.

\subsection{Formulation}
\label{formulation}
Through the discussion above, we wish find a bijection (hopping function, corresponding to { Req. 3.} )
\begin{align}
\begin{array}{rccl}
f:& I\times J & \longrightarrow & I \times J \\
  & (i,j) & \mapsto &(f_1 (i,j),f_2 (i,j))
\end{array}
\end{align}
satisfying:

(i). $f_2 (i_1,j) \neq f_2 (i_2,j)$      for  $i_1 \neq i_2$.     \quad (Solve the half duplex problem, corresponding to { Req. 1.})

(ii). $f_1 (i,j) \neq i$     for any  $(i,j) \in I \times J$.      (Frequency hopping, corresponding to { Req. 2.})

(iii). There exists a set $X$ and a surjection $g: I \times J \longrightarrow X$ such that

\begin{align}
g \circ f=f.    \quad     \mbox{     (Hopping invariant, corresponding to { Req. 4.})}
\end{align}

\section{Solutions of the problem}
Firstly, we derive the solutions in a special case that the column number of pattern is prime in subsection \ref{special case}. Then we generate the conclusions to general case in subsection \ref{general case}.

\subsection{Special case (column number is prime)}
\label{special case}

This subsection is the process to search the solutions. Readers interesting in only results can hop it and go to next subsection.

For simplicity, we suppose that the discovery frame structure is $I=J=\{0,1, \cdots ,p-1  \}$, where $p$ is a prime number,
 and search a bijection $f: I\times J \longrightarrow I \times J$ satisfying above property (i), (ii), (iii) and of affine form.
  i.e
\begin{align}
\begin{array}{rccl}
 f:  I\times J&  \longrightarrow & I \times J \\
    (i,j)& \mapsto &  ((ai+bj+e) \mod p,(ci+dj+f) \mod p)
\end{array}
\end{align}

Because $f$ is a bijection, we know the determinant of the matrix
$\left(\begin{array}{cc}
a & b \\
c & d
\end{array}\right)$
 is coprime to $p$. i.e.
\begin{align}
\label{det}
(ad-bc,p)=1
\end{align}
From (i), we know
\begin{align}
ci_1+dj+f \not \equiv ci_2+dj+f  \mod p
\end{align}
for any $i_1 \not \equiv i_2$ and any $j$, hence
\begin{align}
                               (c, p)=1
\end{align}

\subsubsection{Invariant of degree 1}

 Firstly, we suppose that the hopping invariant in (iii) is a linear map first, i.e. it has the form
\begin{align}
\begin{array}{cll}
					I \times J & \longrightarrow & \mathbb{Z}/p\mathbb{Z}       \\
                              ( i,j) & \mapsto & (xi+yj)  \mod p
\end{array}
\end{align}
where $x$ is not divided by $p$ or $y$ is not divided by $p$. Then we have

\begin{align*}
&(x,y)\left(
\begin{array}{ccc}
a & b & e \\
c & d & f
\end{array}
\right)
\left(
\begin{array}{c}
i\\
j\\
1
\end{array}
\right) \\
\equiv &(x,y,0)\left(
\begin{array}{c}
i\\
j\\
1
\end{array}
\right)   \mod p, \quad \mbox{ for all } i,j
\end{align*}

and hence
\begin{align}
(x,y)\left(
\begin{array}{ccc}
a & b & e \\
c & d & f
\end{array}\right) \equiv (x, y, 0) \mod p
\end{align}
hence
\begin{align}
                        x(a-1,b,e)+y(c,d-1,f) \equiv 0        \mod p.
\end{align}
Because $x$ or $y$ is not divided by $p$ and $(c, p)=1$, we know $x\not \equiv 0  \mod p$. Let $\lambda:=-x^{-1} y \mod p$, then we get
\begin{align}
                    \lambda(c,d-1,f)=(a-1,b,e)
\end{align}
Therefore
\begin{align}
\label{7}
     & a \equiv \lambda c+1 \mod p,  \\
     & b \equiv \lambda (d-1) \mod p, \\
     & e \equiv \lambda f \mod p
\end{align}
From (\ref{det}), we know
\begin{align}
(\lambda c+1)d-\lambda (d-1)c \not \equiv 0  \mod p
\end{align}
hence
\begin{align}
\label{8}
                         d+ \lambda c \not \equiv 0  \mod p
\end{align}
From (ii), we know $(\lambda c+1)i+\lambda (d-1)j+ \lambda f \not \equiv i \mod p$, for any $i$ and $j$. It means the linear
equation
\begin{align}
\label{nosolution}
                          \lambda ci+ \lambda (d-1)j+ \lambda f=0
\end{align}
on finite fields $\mathbb{F}_p$ has not solution, hence

\begin{align}
\lambda c  \equiv 0 \mod p,  \quad \lambda (d-1)  \equiv 0 \mod p, \quad \lambda f \not \equiv 0  \mod p
\end{align}
hence
\begin{align}
& \lambda \not \equiv 0 \mod p, \quad c \equiv 0 \mod p, \\
&  d-1 \equiv 0 \mod p, \quad f \not \equiv 0 \mod p
\end{align}

It contradict that $(c, p)=1$. Hence there is not a bijection $f: I \times J \longrightarrow I \times J$ satisfying
property (a), (b) with a linear invariant for the discovery frame structure $I=J=\mathbb{Z}/p\mathbb{Z}$.

But we see, if
\begin{align}
\lambda c  \not  \equiv 0 \mod p,  \quad \lambda (d-1)  \equiv 0 \mod p, \quad \lambda f  \equiv 0  \mod p
\end{align}
i.e.
\begin{align}
\label{10}
&\lambda \not \equiv 0 \mod p, \quad c \not \equiv 0 \mod p, \\
&  d-1 \equiv 0 \mod p, \quad f \equiv 0 \mod p,
\end{align}
the equation (\ref{nosolution}) has not solution in $I' \times J$, where $I':=I \setminus \{0\}$, and the restriction
\begin{align}
f_{| I'  \times J} : I' \times J \longrightarrow I' \times J
\end{align}
of $f$ on $I' \times J$ is a bijection from $I' \times J$ to itself.

Combine the formula (\ref{7}), (\ref{8}) and (\ref{10}), in fact we got a family $\{ f_{\lambda, c}\}_{\lambda, c \in I', p \nmid   (\lambda c+1) }$ of bijections
\begin{align}
\begin{array}{cccl}
f_{\lambda, c}:& I' \times J & \longrightarrow & I' \times J  \\
& (i, j) & \mapsto & ((\lambda c+1)i \mod p, (ci+j) \mod p)
\end{array}
\end{align}
satisfying the condition (1), (2), (3), where the invariant of $f_{\lambda, c}$ is $(\lambda j-i) \mod p$.

On the basis of above discussion, in fact we get:

\begin{prop}
Let $p$ be a odd prime number, $I':=\{1, 2, \cdots, p-1 \}$, $J:=\{0, 1, \cdots, p-1 \}$. Let $\lambda, c \in I'$ such that $\lambda c+1 \not \equiv 0 \mod p$. Define the map
\begin{align*}
\begin{array}{rccl}
f: & I' \times J & \longrightarrow & I' \times J \\
& (i, j) & \mapsto & ((\lambda c+1)i \mod p, (ci+j) \mod p).
\end{array}
\end{align*}
Then $f$ satisfies the condition (i), (ii), (iii), where $(\lambda j-i) \mod p$ is a hopping invariant.
\end{prop}

\subsubsection{Invariant of degree 2}
 We suppose that the hopping invariant in (3) has degree $2$, i.e. it has the form

\begin{align}
\begin{array}{ccl}
I \times J & \longrightarrow & \mathbb{Z}/p\mathbb{Z} \\
(i, j) & \mapsto & (i, j, 1) \left(
\begin{array}{ccc}
x & u & v \\
u & y & w \\
v & w & z
\end{array}
\right)
\left(
\begin{array}{c}
i\\
j\\
1
\end{array}
\right) \mod p
\end{array}
\end{align}
Hence
\begin{align}
 \left(
\begin{array}{ccc}
a & c & 0 \\
b & d & 0 \\
e & f & 1
\end{array}
\right)
 \left(
\begin{array}{ccc}
x & u & v \\
u & y & w \\
v & w & z
\end{array}
\right)
 \left(
\begin{array}{ccc}
a & b & e \\
c & d & f \\
0 & 0 & 1
\end{array}
\right)  \\
\equiv
 \left(
\begin{array}{ccc}
x & u & v \\
u & y & w \\
v & w & z
\end{array}
\right)  \mod p
\end{align}

From (2), we know that the equation  about $(i,j)$
\begin{align}
(a-1)i+bj+e=0
\end{align}
has not solve $\mathbb{F}_p$. Hence
\begin{align}
a-1 \equiv 0 \mod p, \quad b \equiv 0 \mod p, \quad  e \not \equiv 0 \mod p
\end{align}
Hence
\begin{align}
 \left(
\begin{array}{ccc}
1 & c & 0 \\
0 & d & 0 \\
e & f & 1
\end{array}
\right)
 \left(
\begin{array}{ccc}
x & u & v \\
u & y & w \\
v & w & z
\end{array}
\right)
 \left(
\begin{array}{ccc}
1 & 0 & e \\
c & d & f \\
0 & 0 & 1
\end{array}
\right) \\
\equiv
 \left(
\begin{array}{ccc}
x & u & v \\
u & y & w \\
v & w & z
\end{array}
\right)  \mod p
\end{align}
Then we know that the systems of linear equations
\begin{align}
\left(
\begin{array}{ccccc}
0 & c^2 &  2c & 0 & 0 \\
0 & cd &  d-1 & 0 & 0 \\
e & cf &  ce+f & 0 & c \\
0 & d^2-1  & 0 & 0 & 0 \\
0 & df &  de & 0 & d-1 \\
e^2 & f^2 &  2ef & 2e & 2f
\end{array}
\right)
\left(
\begin{array}{c}
x \\
y \\
u \\
v \\
w
\end{array}
\right)=0
\end{align}
on $\mathbb{F}_p$ has non trivial solutions. Because $(c, p)=1$,  left multiply the invertible matrix
\begin{align}
\left(
\begin{array}{cccccc}
1 & 0 & 0 & 0 & 0 & 0 \\
\frac{1-d}{c^2} & \frac{2}{c} & 0 &0&0&0 \\
0&0&1&0&0&0\\
\frac{(d-1)^2}{c^2} & \frac{2(1-d)}{c} &0&1&0&0\\
0&0&0&0&1&0\\
0&0&0&0&0&1
\end{array}
\right)
\end{align}  to the both side of the equation, we get the equations
\begin{align}
\left(
\begin{array}{ccccc}
      0 &    c^2  &   2c  &     0  &     0\\
      0 &  d + 1  &     0  &     0   &    0\\
      e &    cf & ce + f   &    0  &     c\\
      0   &    0 &      0 &      0   &    0\\
      0  &   df   &  de    &   0  & d - 1\\
    e^2  &   f^2  & 2ef  &   2e   &  2f
\end{array}
\right)
\left(
\begin{array}{c}
x\\
y\\
u\\
v\\
w
\end{array}
\right)=0
\end{align}
on $\mathbb{F}_p$. It is equivalent to the equations
\begin{align}
\left(
\begin{array}{ccccc}
      0 &    c^2  &   2c  &     0  &     0\\
      0 &  d + 1  &     0  &     0   &    0\\
      e &    cf & ce + f   &    0  &     c\\
      0  &   df   &  de    &   0  & d - 1\\
    e^2  &   f^2  & 2ef  &   2e   &  2f
\end{array}
\right)
\left(
\begin{array}{c}
x\\
y\\
u\\
v\\
w
\end{array}
\right)=0
\end{align}
It easy to see that the determined of the coefficient matrix is $4ce^2(d^2-1)$. Because $(c, p)=1, (e, p)=1$ and $p$ is a odd prime, we know that
the equation has non-trivial solution if and only if
\begin{align}
d^2-1 \equiv 0 \mod p.
\end{align}
When $d\equiv 1 \mod p$, we get the solution
\begin{align}
\left(
\begin{array}{c}
x\\
y\\
u\\
v\\
w
\end{array}
\right)=\lambda
\left(
\begin{array}{c}
c\\
0\\
0\\
(f-\frac{ce}{2}) \\
-e
\end{array}
\right) \quad \mbox{ for all } \lambda \in \mathbb{F}_p.
\end{align}
When $d  \equiv -1 \mod p$, we get the solution
\begin{align}
\left(
\begin{array}{c}
x\\
y\\
u\\
v\\
w
\end{array}
\right)=
\lambda \left(
\begin{array}{c}
c^2 \\
4 \\
-2c \\
\frac{c(2f-ce)}{2} \\
ce-2f
\end{array}
\right) \quad \mbox{ for all } \lambda \in \mathbb{F}_p.
\end{align}
When $d \equiv 1 \mod p$, the hopping function has the form:
\begin{align}
\begin{array}{rcl}
f: I \times J & \longrightarrow & I \times J \\
(i, j) & \mapsto & (i+e, ci+j+f),
\end{array}
\end{align}
where $(c, p)=1, (e, p)=1$. It has a hopping invariant
\begin{align}
& (i, j, 1) \left(
\begin{array}{ccc}
c & 0 & f-\frac{ce}{2} \\
0 & 0 &  -e\\
f-\frac{ce}{2} & -e & 0
\end{array}
\right)
\left(
\begin{array}{c}
i\\
j\\
1
\end{array}
\right) \\
& \equiv  ci^2+(2f-ce)i-2ej \mod p
\end{align}
When $d \equiv -1 \mod p$, the hopping function has the form:
\begin{align}
\begin{array}{rcl}
f: I \times J & \longrightarrow & I \times J \\
(i, j) & \mapsto & (i+e, ci-j+f),
\end{array}
\end{align}
where $(c, p)=1, (e, p)=1$. It has a hopping invariant
\begin{align}
& (i, j, 1)
\left(
\begin{array}{ccc}
c^2 & -2c & \frac{c(2f-ce)}{2} \\
-2c & 4 &  ce-2f\\
\frac{c(2f-ce)}{2} & ce-2f & 0
\end{array}
\right)
\left(
\begin{array}{c}
i\\
j\\
1
\end{array}
\right)  \\
\equiv & c^2i^2+4j^2-4cij+c(2f-ce)i+2(ce-2f)j  \mod p
\end{align}
On the basis of above discussion, in fact we get:

\begin{prop}
Let $p$ be a odd prime number, $I=J:=\{0, 1, \cdots, p-1 \}$. Let $c, e \in I \setminus \{0\}$, $f \in I$. Define the map
\begin{align*}
\begin{array}{rccl}
f: & I \times J & \longrightarrow & I \times J \\
& (i, j) & \mapsto & (i+e \mod p, (ci+j+f) \mod p).
\end{array}
\end{align*}
Then $f$ satisfies the condition (i), (ii), (iii), where $(ci^2+(2f-ce)i-2ej) \mod p$ is a hopping invariant.
\end{prop}

\begin{prop}
Let $p$ be a odd prime number, $I=J:=\{0, 1, \cdots, p-1 \}$. Let $c, e \in I \setminus \{0\}$, $f \in I$. Define the map
\begin{align}
\begin{array}{rcl}
f: I \times J & \longrightarrow & I \times J \\
(i, j) & \mapsto & (i+e, ci-j+f),
\end{array}
\end{align}
Then $f$ satisfies the condition (i), (ii), (iii), where $(c^2i^2+4j^2-4cij+c(2f-ce)i+2(ce-2f)j)  \mod p$ is a hopping invariant.
\end{prop}

\subsection{General case}
\label{general case}
We can generalize the conclusions in previous subsection to some more general situation.

\begin{prop}
Let $m$ be a positive odd number, $n$ be a positive integral number divided by $m$. Let $I':=\{1, 2, \cdots, m-1 \}$, $J:=\{0, 1, \cdots, n-1 \}$. Let $u, v$ be two integral numbers such that $(u, m)=1, (u-1, m)=1, (v, m)=1$.
Define the map $f:$
\begin{align*}
\begin{array}{rcl}
I' \times J & \longrightarrow & I' \times J \\
 (i, j) & \mapsto & (ui \mod m, j-vi+v(ui \mod m) \mod n).
\end{array}
\end{align*}
Then $f$ is a bijection satisfying the condition (i), (ii), (iii), where $(j-vi) \mod n $ is a hopping invariant. We call the
hopping patterns determined by such maps "{\bf hopping pattern of type A1}".
\end{prop}
{\bf Proof.} Because $(u, m)=1$, we know $f$ is a bijection. We prove that f satisfies the condition (i), (ii), (iii) as follows:

(i) Suppose that $f_2(i_1, j)=f_2(i_2, j)$, then we have
\begin{align*}
j-vi_1+v(ui \mod m) \equiv j-vi_2+v(ui_2 \mod m)          \mod n.
\end{align*}
Because $n$ is divided by $m$, we have
\begin{align*}
j+v(u-1)i_1 \equiv j-v(u-1)i_2.
\end{align*}
Because $(u-1, m)=1$ and $(v, m)=1$, we know $i_1=i_2$.

(ii) Suppose that $f_1(i, j)=i$, then we have $ui \equiv i \mod m$, and hence $(u-1)i \equiv 0 \mod m$. It contradict
 that $(u-1, m)=1$.

 (iii) It easy to check that the map
\begin{align*}
\begin{array}{ccl}
I' \times J & \longrightarrow & \mathbb{Z}/n\mathbb{Z} \\
(i, j) & \mapsto & (j-vi) \mod n
\end{array}
\end{align*}
satisfies this condition.       \qed

\begin{prop}
Let $m$ be a positive odd number, $n$ be a positive integral number greater than or equal to $m$. Let $I':=\{1, 2, \cdots, m-1 \}$, $J:=\{0, 1, \cdots, n-1 \}$. Define the map
\begin{align*}
\begin{array}{rcl}
f:  I' \times J & \longrightarrow & I' \times J \\
 (i, j) & \mapsto & (2i \mod m, j-i+(2i \mod m) \mod n).
\end{array}
\end{align*}
Then $f$ is a bijection satisfying the condition (i), (ii), (iii),
where $(j-i) \mod n \in \mathbb{Z}/n\mathbb{Z}$ is a hopping invariant.
We call the
hopping patterns determined by such maps "{\bf hopping pattern of type A2}".
\end{prop}
{\bf Proof.} It easy to see that $f$ is a bijection. We prove that f satisfies the condition (i), (ii), (iii) as follows:

(i) We just need to prove that the map
\begin{align*}
\begin{array}{ccl}
\mathbb{Z}/m\mathbb{Z} & \longrightarrow & \mathbb{Z}/n\mathbb{Z}    \\
x & \mapsto & ((2x \mod m)-x) \mod n
\end{array}
\end{align*}
is an injection. Hence we just need to prove that for any $c \in \mathbb{Z}/n\mathbb{Z}$ the equation
\begin{align*}
(2x \mod m) -x \equiv c \mod n
\end{align*}
has at most one solve in $\mathbb{Z}/m\mathbb{Z}$. Hence we just need to prove that for any $c \in [0, m)$, the equation
\begin{align*}
(2x \mod m)  \equiv x + c \mod n
\end{align*}
has at most one solve in $[0, m)$. Hence we just need to prove that for any $c \in [0, m)$, the curve
\begin{align*}
V:\quad  y=2x \mod m \quad (x \in [0, m))
\end{align*}
and the curve
\begin{align*}
V_c: \quad y=(x+c) \mod n    \quad (x \in [0, m))
\end{align*}
have at most one intersection point. We can see it in the following graph.
When $c \in [0, \frac{m}{2})$, $V$ (blue line) and $V_c$ (red line) has one intersection point; when
$c \in [\frac{m}{2}, n-\frac{m}{2}]$, $V$ and $V_c$ (green line) has no intersection point; when $c \in [n-\frac{m}{2}, m)$ (may be empty),
$V$  and $V_c$ (pink line) has one intersection point.

\begin{figure}[H]
\label{VVc}
\includegraphics[width=0.5\textwidth]{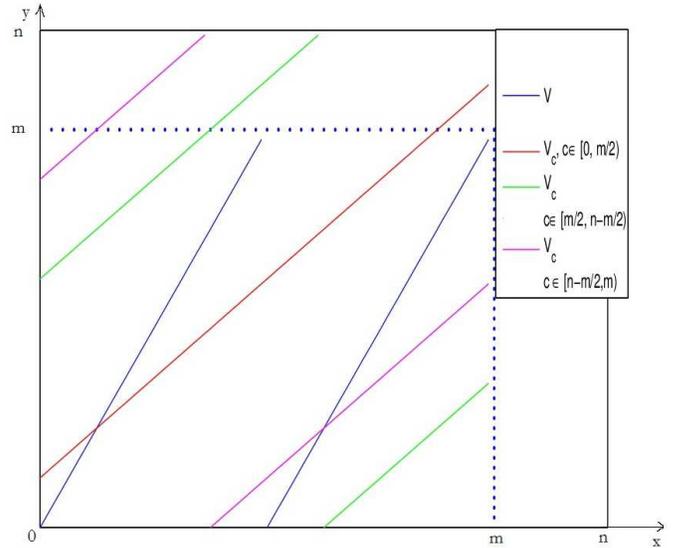}
\caption{Intersection of $V$ and $V_c$}
\end{figure}

(ii) Suppose that $f_1(i, j)=i$, then we have $2i \equiv i \mod m$, and hence $i \equiv 0 \mod m$. It contradict
 that $i \in I'$.

 (iii) It easy to check that the map
\begin{align*}
\begin{array}{ccl}
I' \times J & \longrightarrow & \mathbb{Z}/n\mathbb{Z} \\
(i, j) & \mapsto & (j-i) \mod n
\end{array}
\end{align*}
satisfies this condition.       \qed

\begin{prop}
Let $m$ be a positive odd number, $n$ be a positive integral number divided by $m$.
Let $I:=\{0, 1, \cdots, m-1 \}$, $J:=\{0, 1, \cdots, n-1 \}$. Let $e$ be an integral number not divided by $m$, $c$ be an integral number coprime to $m$, $f$ be an integral number.
Define the map
\begin{align*}
\begin{array}{rccl}
f: & I \times J & \longrightarrow & I \times J \\
& (i, j) & \mapsto & (i+e \mod m, (ci+j+f) \mod n).
\end{array}
\end{align*}
Then $f$ is a bijection satisfying the condition (i), (ii), (iii), where $(ci^2+(2f-ce)i-2ej) \mod m$ is a hopping invariant.
We call the
hopping patterns determined by such maps "{\bf hopping pattern of type B1}".
\end{prop}

{\bf Proof.} It easy to see that $f$ is a bijection. We prove that f satisfies the condition (i), (ii), (iii) as follows:

(i) Suppose that $f_2(i_1, j)=f_2(i_2, j)$, then we have
\begin{align*}
ci_1+j+f \equiv ci_2+j+f          \mod n.
\end{align*}
and hence $ci_1 \equiv ci_i \mod m$ because $n$ is divided by $m$. Therefore we have $i_1 \equiv i_2 \mod m$, because $c$
is coprime to $m$.

(ii) It is trivial, because $e$ is not divided by $m$.

 (iii) It easy to check that the map
\begin{align*}
\begin{array}{ccl}
I \times J & \longrightarrow & \mathbb{Z}/m\mathbb{Z} \\
(i, j) & \mapsto & (ci^2+(2f-ce)i-2ej) \mod m
\end{array}
\end{align*}
satisfies this condition.       \qed

\begin{prop}
Let $m$ be a positive odd number, $n$ be a positive integral number divided by $m$.
Let $I:=\{0, 1, \cdots, m-1 \}$, $J:=\{0, 1, \cdots, n-1 \}$. Let $e$ be an integral number not divided by $m$, $c$ be an integral number coprime to $m$, $f$ be an integral number.
Define the map
\begin{align}
\begin{array}{rcl}
f: I \times J & \longrightarrow & I \times J \\
(i, j) & \mapsto & (i+e, ci-j+f),
\end{array}
\end{align}
Then $f$ satisfies the condition (i), (ii), (iii), where $(c^2i^2+4j^2-4cij+c(2f-ce)i+2(ce-2f)j)  \mod n$ is
a hopping invariant. We call the
hopping patterns determined by such maps "{\bf hopping pattern of type B2}".
\end{prop}

{\bf Proof.} It is similar to the proof of above proposition.     \qed

Based on the discussion above, we have obtained four kinds of patterns with hopping invariant, and for further detailed understanding of theses patterns, we compare them with these from QC as below.

\begin{table}[h] \scriptsize
\begin{tabular}[t]{|l|c|c|c|c|}
\hline
pattern & time  & frequency & independent & have hopping  \\
&hopping & hopping & of $t$ & invariant\\
\hline
QC($c \equiv 0 \mod m$) & Y & N & Y & Y  \\
\hline
QC($c \not \equiv 0 \mod m$) & Y & Y & N & N \\
\hline
type A1 & Y & Y & Y & Y \\
\hline
type A2 & Y & Y & Y & Y \\
\hline
type B1 & Y & Y & Y & Y \\
\hline
type B2 & Y & Y & Y & Y \\
\hline
\end{tabular}
\end{table}

From this table, we can see that the hopping patterns of type A1, A2, B1, B2 have all the advantage we hoping: time hopping, frequency hopping, independent of $t$ and hopping invariant.

\section{Conclusion}
In this paper, we discover and utilize the property "hopping invariant" to design several hopping patterns to enable high efficient D2D discovery procedure. Besides, we consider the basic requirement of hopping pattern and the specific system constraints, which makes our schemes more robust and suitable to be implemented in practical systems such as LTE. Furthermore, the application of "hopping invariant" leads to efficient information delivery as no extra system resources are needed and the receiver complexity can be reduced significantly since the embedded connection between resource location and the hopping invariant. After comparison, it is shown that our schemes outperform other referred ones in terms of frequency hopping, independence of frame number and hopping invariant.

\end{document}